\documentclass[usenatbib,letterpaper]{mn2e}
\usepackage{mathptmx}
\usepackage{fixltx2e}
\usepackage{epsfig}
\usepackage{natbib}		

\newcommand{\Gyr}{\,\mbox{Gyr}}
\newcommand{\Mpc}{\,\mbox{Mpc}}
\newcommand{\kpc}{\,\mbox{kpc}}
\newcommand{\kms}{\,\mbox{km}\,\mbox{s}^{-1}}

\newcommand{\msun}{\,M_{\sun}}
\newcommand{\degree}{^\circ}

\newcommand{\tsim}{\sim\!}
\newcommand{\ea}{et al.~}
\newcommand{\nbody}{$N$-body}

\newcommand{\ltrsim}{\la}

\newcommand{\rsh}{r_{sh}}
\newcommand{\vsh}{v_{sh}}
\newcommand{\vlos}{v_{los}}
\newcommand{\half}{^{1/2}}
\newcommand{\feh}{[\mathrm{Fe}/\mathrm{H}]}
\newcommand{\softfont}[1]{\textsc{#1}}
\title[A spectroscopic survey of Andromeda's Western Shelf] 
  {A spectroscopic survey of Andromeda's Western Shelf}
\author[M. A. Fardal et al.]       
{Mark A. Fardal$^1$\thanks{E-mail: fardal@astro.umass.edu}, 
  Puragra Guhathakurta$^2$, 
  Karoline M. Gilbert$^{3,11}$, 
  Erik J. Tollerud$^4$, \newauthor
  Jason S. Kalirai$^{5,6}$,
  Mikito Tanaka$^{7,8}$, 
  Rachael Beaton$^9$,
  Masashi Chiba$^7$,\newauthor
  Yutaka Komiyama$^{10}$, 
  Masanori Iye$^{10}$\\
 $^1$Dept.\ of Astronomy, University of Massachusetts, 
     Amherst, MA 01003, USA \\
 $^2$UCO/Lick Observatory, Dept.\ of Astronomy \& Astrophysics,
     Univ. of California, 1156 High St., Santa Cruz, CA 95064, USA \\
 $^3$Department of Astronomy, University of Washington, Box 351580, 
     Seattle, WA 98195, USA \\
 $^4$Center for Cosmology, Department of Physics and Astronomy, 4129 Frederick 
     Reines Hall, University of California, Irvine, CA 92697, USA \\
 $^5$Space Telescope Science Institute, Baltimore, MD 21218, USA \\
 $^6$Center for Astrophysical Sciences, Johns Hopkins University, 
     Baltimore, MD 21218, USA\\
 $^7$Astronomical Institute, Tohoku University, Aoba-ku, Sendai
     980-8578, Japan \\
 $^8$University of Tokyo, 7-3-1 Hongo, Bunkyo-ku, Tokyo 113-0033, Japan\\
 $^9$Department of Astronomy, University of Virginia, P.O. Box
     3818, Charlottesville, VA 22903, USA\\
 $^{10}$National Astronomical Observatory of Japan, 
     2-21-1 Osawa, Mitaka, Tokyo 181-8588, Japan\\
 $^{11}$Hubble Fellow
}  
\date{\today}
\pagerange{\pageref{firstpage}--\pageref{lastpage}} \pubyear{2012} 
\begin{document}
\maketitle  
\label{firstpage}
\begin{abstract}
The Andromeda galaxy (M31) shows many tidal features in its halo,
including the Giant Southern Stream (GSS) and a sharp ledge in surface
density on its western side (the W Shelf).  Using DEIMOS on the Keck
telescope, we obtain radial velocities of M31's giant stars along its
NW minor axis, in a radial range covering the W Shelf and extending
beyond its edge.  In the space of velocity versus radius, the sample
shows the wedge pattern expected from a radial shell, which is
detected clearly here for the first time.  This confirms predictions
from an earlier model of formation of the GSS, which proposed that the
W Shelf is a shell from the third orbital wrap of the same tidal
debris stream that produces the GSS, with the main body of the
progenitor lying in the second wrap.  We calculate the 
distortions in the shelf wedge pattern expected from its outward
expansion and angular momentum, and show that these 
effects
are echoed in the data.  In addition, a hot, relatively smooth
spheroid population is clearly present.  We construct a
bulge-disk-halo \nbody\ model that agrees with surface brightness and
kinematic constraints, and combine it with a simulation of the GSS.
From the contrasting kinematic signatures of the hot spheroid and
shelf components, we decompose 
the observed stellar metallicity distribution into contributions
from each component using a non-parametric mixture model.  
The shelf component's metallicity distribution
matches previous observations of the GSS superbly, further
strengthening the evidence they are connected and bolstering the case
for a massive progenitor of this stream.
\end{abstract}
\begin{keywords}
galaxies: M31 -- galaxies: interactions -- galaxies: kinematics and dynamics
\end{keywords}

\section{INTRODUCTION}
\label{sec.intro}
The Andromeda galaxy (M31) lies close enough for us to resolve its
individual stars, at a position relatively unobscured by the Milky Way
(MW) disk.  Thus it represents perhaps our best opportunity to examine the
complete structure of a large spiral galaxy in great detail.  The
region that we might term M31's inner halo or outer disk appears quite
irregular in resolved-star maps
\citep{ibata01,ferguson02,irwin05,ibata07,alan09,tanaka10}.  
These maps show numerous
substructures, some of which have been catalogued and
named, although stars are distributed continuously throughout M31's halo 
out to at least 150~kpc.

\begin{figure}
\centering
\includegraphics[width=6.59cm]{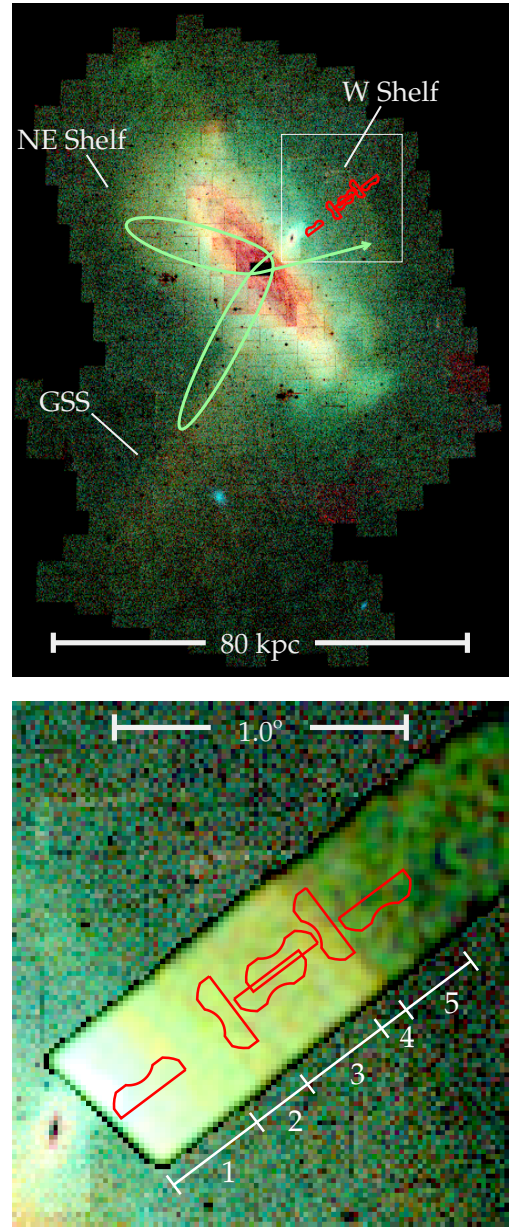}
\caption{
\label{fig.masks}
Top: star-count map from the INT survey of M31 
\citep[][image created by M.~Irwin]{irwin05}, with our 
Keck/DEIMOS spectroscopic multislit masks indicated in red.
Blue, green, and red channels indicate the
density of stars in separate color ranges on the RGB.
The red channel contains the reddest, most metal-rich RGB stars.
M31's disk is
apparent in red, and the main tidal features discussed in the text are
indicated by labels.  The W Shelf in particular is just visible as a
faint but sharp edge to the NW of M31. The box shows the area covered
in the bottom panel.  The looping arrow indicates the progenitor's path
in our model.  
Bottom: enlargement of the region around our
spectroscopic masks.  Here a matched-filter map from the 
Subaru/SuprimeCam imaging survey
\citep{tanaka10} is overlaid, forming a diagonal stripe from 
lower left to upper right.
The radial zones 1--5 used later are indicated.}
\end{figure}

Pencil-beam spectroscopy and photometry brings some order to this
complicated scene.  \citet{ibata05} argues much of the structure is an
outer extension of M31's disk, with a clumpy structure and hot or thick
disk kinematics.  It also seems much of the structure is due to M31's
giant southern stream or GSS \citep{ibata01}.  \citet{ferguson02}
first suggested a connection between the GSS and the ``NE Shelf''
feature extending from the NE side of the M31 disk 
(see Fig.~\ref{fig.masks}),
which appears to have a red population of red giant branch
(RGB) stars and thus is inferred to be metal-rich.  \citet[][hereafter
F07]{fardal07} pointed out a similar shelf on the western side of M31
(W Shelf), which they inferred to be a third wrap of the stream using
an \nbody\ simulation of the stream's formation.  \citet{gilbert07} found
a structure through RGB star kinematics on the southeast minor axis
(the ``SE Shelf'') and identified it as the {\it fourth} wrap of this
stream.  Indeed, \citet{richardson08} found most of the dense
structures in the inner halo examined by their deep 
Hubble Space Telescope (HST)
observations could be classified as either ``disk-like'' or
``stream-like'', with the positions of the ``stream-like'' fields
matching well to the F07 model.

These two components cannot be the whole story, however.  Spectroscopy 
also offers evidence of a smoother hot halo component
\citep{raja05,raja06,chapman06,kalirai06b,gilbert07}.  There appears to 
be an upward break in the surface brightness profile around 20--$30 \kpc$
\citep{irwin05,raja05,ibata07} and a decrease in the metallicity in
about the same place \citep{kalirai06b,koch08}.  
It is much disputed whether the ``inner halo'' region from 10--$20
\kpc$ is dominated by 
a halo component \citep{pritchet94,durrell01}
the bulge \citep{raja05,kalirai06b},
the disk \citep{ibata07,courteau11}, 
or by a mix of these plus cold substructures \citep{gilbert07}.
Other kinematic anomalies such as the ``second stream'' detected by
\citet{kalirai06a} and \citet{gilbert09} currently lack an
explanation.  One might also expect tidal features from M32 or NGC 205
(M110), due to their location near M31's center and distorted
isophotes \citep{choi02}.

To better understand this complex structure, it is important
to deduce how many components there are and determine for each its
distribution in space, velocity, stellar age, and metallicity.  A
key ingredient in this will be a firm determination of the orbit of 
the GSS and the spatial coverage of its debris.  The F07 model
matches many features of the morphology and kinematics, but it is not
the only suggestion for the orbit of the stream
\citep{ibata01,merrett03,fardal06,hammer10}.  
A crucial test of the model comes in the W
shelf region.  The shelf stars are expected to exhibit a
distinctive kinematic pattern in this model.  F07 presented
some kinematic evidence for this model from planetary nebulae
and RGB stars, but this was more suggestive than conclusive.

In this paper we present new spectroscopy of the resolved stellar
population in the W Shelf
region that shows clear evidence for distinct shelf and hot spheroid
components.  This kinematic data agrees very well with the
prediction from an updated version of the F07 model,
which was run before the observations were reduced.  \S 2 provides
details on the observations and our \nbody\ model. \S 3 analyzes the
observations, first showing (\S 3.1) the distributions of the stars in
position, velocity, and metallicity space.  
We then use the \nbody\ model (\S 3.2) to separate the shelf and
hot halo components in the observations, and thereby infer their
distinct metallicity distributions.
\S 4 concludes the paper by connecting our results to previous
results on M31's GSS and other inner halo populations, and
placing our results in the broader context of halo structure in 
other galaxies.  
An Appendix provides a discussion of shell kinematics, including a
formula for the velocity envelope in the generically correct case of
an expanding shell.
\section{METHODS}
\label{sec.methods}

\subsection{Observational data}
\label{sec.obs}

\begin{table*}
\begin{minipage}{180mm}
\caption{Spectroscopic observations}
\begin{tabular}{lcrrrrlllrrr}  
\hline
\multicolumn{1}{c}{} & 
\multicolumn{1}{c}{Radial} & 
\multicolumn{1}{c}{Radius} & 
\multicolumn{2}{c}{Pointing center:} &
\multicolumn{1}{c}{PA} & 
\multicolumn{1}{c}{Observation} & 
\multicolumn{1}{c}{Exposure}  & 
\multicolumn{1}{c}{Seeing}  & 
\multicolumn{1}{c}{Science} & 
\multicolumn{1}{c}{Confirmed} & 
\multicolumn{1}{c}{M31} \\
\multicolumn{1}{c}{Mask} & 
\multicolumn{1}{c}{Zone} & 
\multicolumn{1}{c}{(kpc)} & 
\multicolumn{1}{c}{$\alpha_{\rm J2000}$} &
\multicolumn{1}{c}{$\delta_{\rm J2000}$} & 
\multicolumn{1}{c}{($\degree$)} & 
\multicolumn{1}{c}{Date (UT)} & 
\multicolumn{1}{c}{Time (s)} &  
\multicolumn{1}{c}{('')} & 
\multicolumn{1}{c}{Targets$^a$}  & 
\multicolumn{1}{c}{Stars$^b$}  & 
\multicolumn{1}{c}{Stars$^c$}  \\ 
\hline
NW1V\_1  & 1 &  13.1 &  0:38:36.33 &  41:50:11.8 & $ 127  $ & 2009 Aug 22  & 3600 &  0.5--0.8  &         279 &   56 &  46 \\
NW1dV    & 2 &  18.1 &  0:37:07.94 &  42:05:35.8 & $ -143 $ & 2009 Aug 21  & 3600 &  0.6--0.7  &	 213 &  102 &  78 \\
NW2V\_1  & 3 &  20.6 &  0:36:20.72 &  42:12:08.5 & $ -53  $ & 2009 Aug 21  & 3000 &  0.5--0.6  &	 213 &   89 &  62 \\
NW2V\_2  & 3 &  21.6 &  0:36:07.90 &  42:15:48.4 & $ 127  $ & 2009 Aug 21  & 3000 &  0.5--0.7  &	 208 &  100 &  62 \\
NW2dV    & 4 &  24.2 &  0:35:18.87 &  42:23:22.8 & $ -143 $ & 2009 Aug 22  & 3300 &  0.5--0.6  &	 201 &  124 &  90 \\
NW3V\_1  & 5 &  26.8 &  0:34:22.67 &  42:28:13.8 & $ -53  $ & 2009 Aug 21  & 3000 &  0.5       &	 193 &   76 &  25 \\
\hline
\end{tabular}

\medskip
$^a$Total number of slits on the mask.\\
$^b$Objects with secure radial velocities 
identified as either MW dwarfs or M31 giants.\\
$^c$Objects with secure radial velocities identified as M31 giants.
\label{table.masks}
\end{minipage}
\end{table*}

The data in this paper was obtained as part of the Spectroscopic and
Photometric Landscape of Andromeda's Stellar Halo (SPLASH) survey.
Our spectroscopic survey area is embedded within the imaging survey 
of \citet{tanaka10}, conducted with Suprime-Cam on the Subaru Telescope.
That survey had the goal of quantifying the surface brightness profile
and metallicity of the M31 halo and detecting streams and other 
substructure, by means of deep color-magnitude diagrams (CMDs) obtained mostly 
along the SE and NW minor axes of M31.
The SE minor axis had earlier been studied intensively by means of wide-field
photometry of individual stars, extremely deep CMDs
obtained with HST, and resolved-star spectroscopy
\citep{pritchet94,durrell01,brown03,irwin05,brown06b,brown07,brown08,ibata07,gilbert07}.
In part because of increasing MW foreground to the north,
the NW minor axis had received much less attention.
Along the NW minor axis, the Subaru survey consists of a series 
of partially overlapping images that form a continuous stripe 
in the range $11 \kpc \ltrsim R \ltrsim 104 \kpc$.
The survey used $V$ and $I$ filters 
on the Johnson-Cousins Vega-based magnitude system 
and probed down to 50\% completeness limits of $V=25.7$, $I=24.9$.

Our spectroscopic survey leverages the earlier imaging
survey to obtain kinematics of M31's inner spheroid region,
in particular covering the W Shelf feature that is prominent
in Fig.~\ref{fig.masks}.  The blue, green, and red channels 
of the Subaru map are constructed 
from explicit cuts in metallicity with boundaries in $\feh$ of 
$-2.31$, $-1.71$, $-0.71$, and $0.0$, and due to their different
input data and construction do not match the larger-scale
Isaac Newton Telescope (INT) map exactly.
Nevertheless, both maps show the W Shelf clearly as a 
red, metal-rich feature with a sharp edge,
which is more easily seen in the narrow Subaru portion
in the lower panel.
The observation pattern of the Subaru survey involved small
regions where adjacent Suprime-Cam pointings fields overlapped, and the
resulting source catalog is broken up into regions with either
single or double coverage.
We orient the DEIMOS masks parallel to M31's minor axis, 
except within the overlap regions, where we orient the
masks perpendicular to the minor axis to fit inside the 
double coverage area.
The six spectroscopic masks are described in Table~\ref{table.masks} 
and shown in Fig.~\ref{fig.masks}.
Because two of the masks nearly overlap, these six masks result 
in five separated radial ``zones'', labeled 1--5 from innermost 
to outermost in the discussion below.
Comparing to the maps from the INT and Subaru imaging surveys 
shown in Fig.~\ref{fig.masks},
we can see that zones 1--4 lie within the boundary of the W~Shelf
feature and zone 5 just outside of it.  

The spectroscopic targets were chosen from the catalog of sources
obtained from the Subaru survey of M31 \citep{tanaka10}.
This catalog was produced using PSF-fitting photometry using 
\softfont{DAOPHOT-II}.
Extended objects (mostly galaxies, as well as some blended stars) 
were removed from the target
sample based on a set of \softfont{DAOPHOT-II} morphological parameters.  
Details of this procedure are in \citet{tanaka10}.

At these relatively high surface brightness levels in Andromeda's halo,
we expect from previous experience that
most of the stellar targets 
are M31 stars, at least below the tip of the RGB at M31's 
distance at $I = 20.5$.  
For simplicity, then, our target selection uses 
only the $I$ magnitude, which roughly measures the flux in the
wavelength range of our spectra.
We choose targets in the range $20.5 < I < 23.0$.
We assign a priority to each target which decreases linearly
with magnitude by a factor of two over this range.
Our mask design software then chooses a slit pattern with 
the aim of maximizing the sum of priorities over the mask using 
a minimum slit length criterion.
This produces a bias toward selection of the brighter stars,
though stars are selected in significant numbers within the
entire magnitude range specified.

Observations were obtained with DEIMOS at the Keck II telescope
in the fall of 2009 in good seeing conditions.
The observational setup was similar to that in \citet{gilbert07}.
We observed 3 sub-exposures for each of the six masks, with
a total observing time of 45 to 60 minutes.
Details of the observations are listed in Table~\ref{table.masks}.
We used the DEIMOS spectrograph with a $1200 \, \mbox{mm}^{-1}$ grating.
The central wavelength was 7800 \AA, with a range of about
6450--9150~\AA\ (depending somewhat on the slit location on the mask).
The dispersion was 0.33~\AA~$\mbox{pixel}^{-1}$, and the 
spectral resolution about 1.3 \AA\ FWHM.

The spectra were reduced and analyzed using a modified version of the 
\softfont{SPEC2D} and \softfont{SPEC1D} software developed by the DEEP2 team 
at the University of California, Berkeley\footnote{{\tt
http://astron.berkeley.edu/$\sim$cooper/deep/spec2d/primer.html},
\newline\indent
{\tt http://astron.berkeley.edu/$\sim$cooper/deep/spec1d/primer.html}
}. These
routines perform standard spectral reduction steps, including flat-fielding, 
night-sky emission line removal, and 
extraction of one-dimensional spectra from the two-dimensional spectra.
Reduced one-dimensional spectra were cross-correlated with a library of 
template
stellar spectra to determine the radial velocity of the object.  Each 
spectrum was visually inspected and assigned a quality code
based on the number and quality of absorption lines.  Spectra with at least
two spectral features (even if one of them is marginal) were 
considered to have secure velocity measurements.
The position of the atmospheric A-band in the
observed spectrum was used to correct the observed velocity for
imperfect centering of the star in the slit 
\citep{simon07,sohn07}.
A heliocentric correction was applied to the measured 
velocities based on the sky position of each mask and the date 
and time of the observation.
More details of these procedures are provided in earlier papers,
including \citet{raja06} and \citet{gilbert07}.

On several of the masks, a significant fraction of the slits 
failed to yield acceptable spectra.  Most of these failures were
localized by position, so that broad swaths of the mask area are missing
any velocity measurements.
These failures were caused by temporary problems with the CCD readout 
electronics and possible problems with the calibration data.
We are confident that the rate of such failures has no intrinsic 
dependence on velocity, and thus introduces no bias into our kinematic
analysis below.

The median velocity error for the stars presented in 
this paper is $3 \kms$, and the maximum is $13 \kms$.
Errors are estimated from the cross-correlation profile,
including a modest scaling factor that is empirically
calibrated by previous duplicate measurements of individual 
stars on overlapping DEIMOS masks.
We ignore these small velocity errors in the analysis
below, as the intrinsic widths of the velocity distributions 
we are interested in are much larger.

For our metallicity estimates we use photometric rather
than spectroscopic methods, since for our combination of 
deep photometry and moderate $S/N$ spectroscopy 
we judge the former to be more reliable.  We estimate metallicities 
from the $V$ and $I$ magnitudes using the procedure described in 
\citet{gilbert07} (rather than that in \citealp{tanaka10}),
for consistency with the dwarf-giant separation algorithm.
The $\feh$ values are based on a comparison of the star's position within the 
CMD to a finely spaced grid of theoretical 12 Gyr, [$\alpha$/Fe]~$=0$ 
stellar isochrones \citep{kalirai06b,vdb06} adjusted to the distance of M31.
Metallicities of the MW stars as assigned by this algorithm 
are therefore not meaningful.

We classify the stars in the sample using the likelihood method of 
\citet{gilbert06}. 
We use the following criteria:
\begin{itemize}
\item[-] The measured equivalent width of the Na~I doublet at 8190 \AA\
combined with the $V-I$ color of the star.
\item[-] The strength of the Ca~II triplet absorption lines, as measured
by comparison of the star's photometric (CMD-based) vs.\ spectroscopic 
(Ca triplet-based) metallicity estimates.
\item[-] The position of the star in an $(V,I)$ CMD
with respect to theoretical RGB isochrones.
\item[-] The radial velocity of the star.
\end{itemize}
Gilbert \ea showed that this combination of diagnostics can yield
a reasonably reliable separation of MW from M31 stars.
The method of \citet{gilbert06} can also incorporate additional
criteria such as photometry in the surface-gravity-sensitive DDO51 
band, but this was not available for our sample.
We flag stars with a M31 membership probability of greater
than 50\% and CMD locations reasonably close to the RGB as M31 stars
(classes 1--3 in the terminology of \citealt{gilbert06}).  
The membership probability is quite
bimodal, and over the sample of identified M31 stars, the 
mean membership probability is 94\%.
A few misclassifications are inevitable, particularly in the 
higher-velocity part of the sample ($v \sim -100 \kms$)
where the M31 and MW velocity distributions overlap
(see \citealp{gilbert07} for a full discussion).  
Our final sample contains 547 stars, of which 363 are flagged as
M31 giants and 184 as MW dwarfs.

\subsection{Simulation}
\label{sec.sim}
The \nbody\ simulation we use in this paper is a slightly updated
version of the simulation in F07.  From our ongoing work with
automated fitting of the GSS (Fardal et al., in preparation), we selected a
set of parameters that offered good agreement with a set of
observational constraints.  The orbital initial conditions in the sky
coordinate system (see \citealp{fardal06}) are $x = -24.0 \kpc$, $y =
0.09 \kpc$, $z = 44.1 \kpc$, $v_x = 29.1 \kms$, $v_y = 8.7 \kms$, $v_z
= -77.6\kms$, and the model was run for $0.972 \Gyr$ to reach its
present point.  
The input satellite is a Plummer model with mass $M_s = 3.09 \times
10^9 \msun$ and scale length $r_s = 1.16 \kpc$, created with the
\softfont{ZENO} library.  
We take the mass to be purely stellar here.
We reason that much of the dark halo mass will have
been stripped off in earlier pericentric passages, as is 
typically found in cosmological simulations \citep{libeskind11}
and analytic models \citep{watson12} of satellite accretion.
The initial mass indicates a relatively massive galaxy rather than
a dark-matter-dominated dwarf.
\citet{mori08}
have considered a GSS progenitor with a massive halo, finding that a
total satellite mass of $\ltrsim 5 \times 10^9 \msun$ is required to avoid
perturbing M31's disk too much.  
(The effects of dark matter in the progenitor will be
considered in more depth in Fardal et al., in preparation, where
we find that the moderate amounts of dark matter allowed by
various constraints do not radically change the nature of the
simulations.)  

The progenitor initially follows 
the trajectory shown in Fig.~\ref{fig.masks} (which assumes a test-particle
orbit), and though its tidal debris is significantly extended it generally
follows a similar trajectory.  This makes the GSS the first
wrap of the orbit since the initial pericenter, the NE shelf the
second, and the W Shelf the third.  The W Shelf grows outwards 
over the last 300--400 Myr from M31's center to its current size.  
For convenience we refer throughout to all the particles from this
simulation as ``GSS'' material, except when explicitly distinguishing 
different radial wraps.

We base our fixed gravitational potential model on \citet{geehan06}.
We use a Hernquist bulge with 
$r_b = 0.61 \kpc$ and 
$M_b = 3.39 \times 10^{10} \msun$,
a Miyamoto-Nagai disk (replacing our earlier exponential disk 
for computational speed) with 
$a_d = 5.94 \kpc$,
$b_d = 0.4 \kpc$, and
$M_d = 8.07 \times 10^{10} \msun \kpc^{-2}$,
and a Navarro-Frenk-White (NFW) halo \citep{navarro96} with 
$r_h = 24.2 \kpc$,
$\rho_h \equiv \delta_c \rho_c = 5.85 \times 10^6 \msun \kpc^{-3}$,
and $M(r<r_h) =  2.02 \times 10^{11} \msun$.
Our model is qualitatively quite similar to that in F07, but better
fits the velocity in the GSS itself and the radii of the NE and W shelves.
The run was conducted with the code PKDGRAV, in advance of the 
observational reductions.\footnote{
An animation and data comparisons for this run can be found in
the first author's presentation at
{\tt www.ari.uni-heidelberg.de/meetings/milkyway2009/talks} and at
{\tt www.astro.umass.edu/\textasciitilde{}fardal/movie\_sphr\_aug09.gif} .}

\begin{figure}
\includegraphics[bb=34 11 479 483,width=8.5cm]{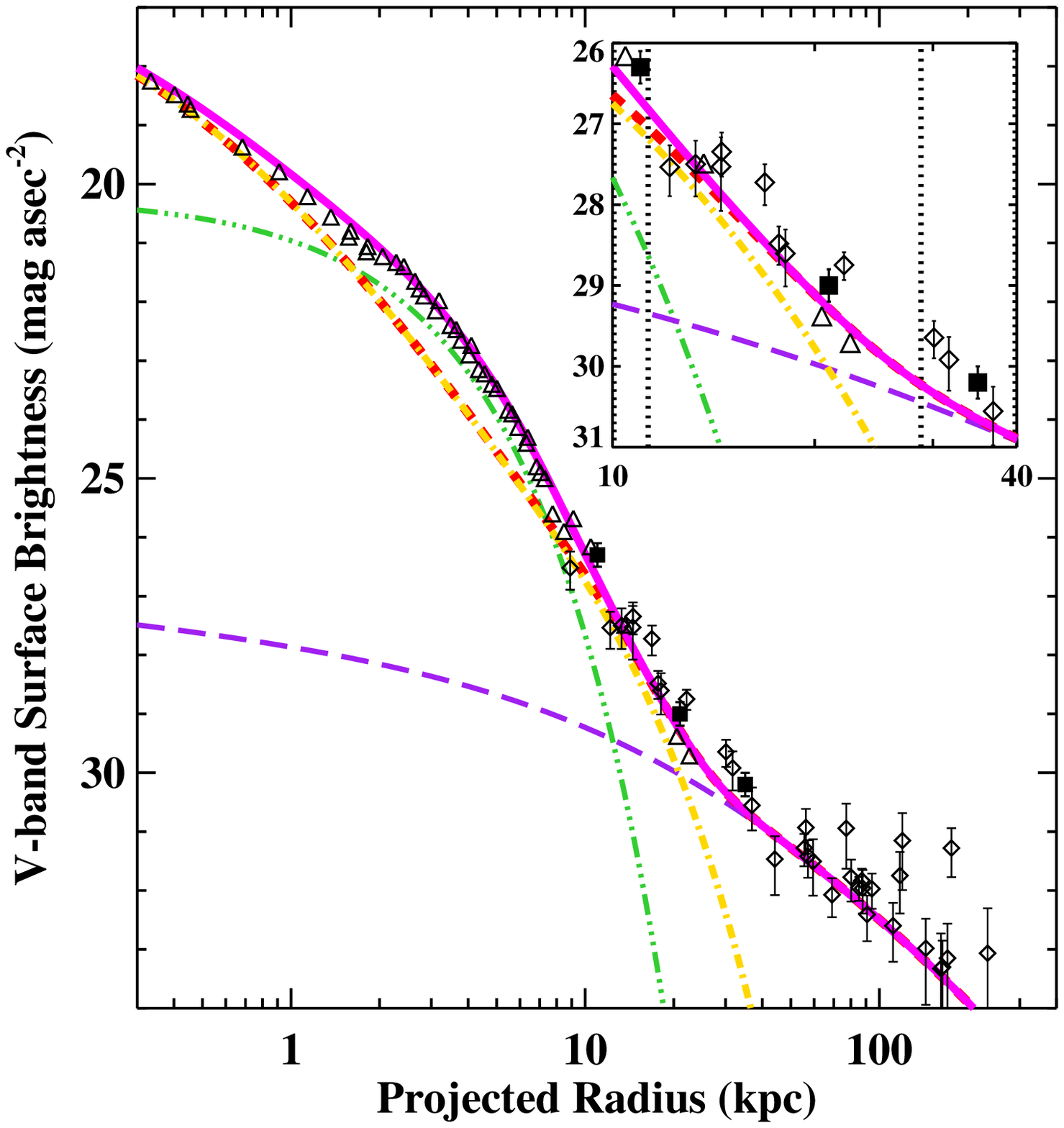}
\caption{
\label{fig.sb}
$V$-band surface brightness distribution of our model, compared to
data.
Solid magenta: total model surface brightess.  
Dot-dashed yellow: bulge.  
Triple-dot-dashed green: disk.  
Long-dashed purple: halo.
Dashed red: bulge and halo combined (or ``spheroid'').  
Solid squares: Brown \ea (2003, 2007, 2008) photometric fields. 
Open triangles: \citet{pritchet94} photometric fields.
Open diamonds: Gilbert \ea (in preparation) spectroscopic fields.
All except the Gilbert points are taken from the minor axis.
The inset expands the region studied in this paper; dashed vertical 
lines show the complete radial range of the target stars.
}
\end{figure}

We supplement this GSS simulation with an \nbody\ realization of M31
itself.  We draw parameters for the bulge from the model just listed.
However, for better agreement with the outer surface brightness profile 
of M31, we truncate the bulge density starting at $11 \kpc$ using the 
default \softfont{ZENO} truncation scheme, which assumes continuous values 
and slopes at the truncation point and thereafter adopts a form 
$\rho(r) \propto r^{-2} \exp(-r/r_\ast)$.  
For the disk we use a true exponential disk as in the \citet{geehan06}
model, with $M_d = 7.34 \times 10^{10} \msun \kpc^{-2}$ and $r_d = 5.4
\kpc$, which is similar to the Miyamoto disk used in the simulation in
terms of its dynamical effects.  We add a stellar halo component not
present in the \citet{geehan06} model, postulating that it follows the
dark halo component and thus has $r_h = 24.2 \kpc$.  The mass is
normalized to match surface brightness measurements and kinematic
samples as described below.  We set the
characteristic density $\rho_0 = \delta_c \rho_c = 1.30 \times 10^4
\msun \kpc^{-3}$, or $M(r<r_h) = 4.50 \times 10^8 \msun$.  As with the
input satellite, \softfont{ZENO} generates the spheroid component particle
velocities from the full distribution function generated by inversion
of an Abel integral, to ensure equilibrium with the potential of the
baryonic components and dark halo.  Although we do not evolve the
\nbody\ snapshot, from previous test simulations we know it is very close 
to equilibrium.

For the combined model, the virial radius (here defined as the radius
where the mean enclosed density is 200 times the critical density,
assuming $H_0 = 71 \kms \Mpc^{-1}$) is $r_{200} = 242 \kpc$ and the
corresponding virial mass is $M_{200} = 1.66 \times 10^{12} \msun$.
(For a virial threshold of 100, these quantities would be $319
\kpc$ and $1.91 \times 10^{12} \msun$ respectively.)  The mass within
$125 \kpc$, sometimes used as a benchmark of M31's halo mass, 
is $1.13 \times 10^{12} \msun$.  The mass of the stellar
halo component within $r_{200}$ is $1.2 \times 10^9 \msun$,
more than a factor of ten lower than the bulge mass.

The surface brightness distribution produced by these
components is shown in Fig.~\ref{fig.sb}.  Here
we assume {\it observed} (not dust-corrected) $M/L_V$ values of 
5.6 for the bulge and 5.0 for the disk, based on estimates of their 
$M/L_R$ and $V-R$ color in \citet{geehan06}.  
For the halo we assume a lower $M/L_V = 2.5$, which is reasonable
for old stellar isochrones of the low metallicity ($\feh < -1.0$)
thought to be prevalent in the halo \citep{kalirai06b,koch08}.
The smaller $M/L_V$ of the halo compared to the bulge is justifiable
based on the lower metallicity and extinction in the halo.  
Of course, these $M/L_V$ values are fairly uncertain, and adopting 
a single $M/L_V$ for each component is itself a gross oversimplification.

There have been many observational studies of M31's surface
brightness, most of which have concentrated on the SE minor axis.  We
include three datasets here: 
the photometric survey of Pritchet and van den Bergh \citep{pritchet94}, 
the spectroscopic survey of Gilbert \ea (in preparation),
and three points from deep HST surveys 
\citet{brown03,brown07,brown08}.
The Gilbert \ea values are calculated after removal of obvious substructure,
and have been normalized to the surface brightness in the HST deep fields.
The HST values in turn were calculated by summing the individual 
stellar luminosities using data products in \citealp{brown09}.
This method should be more accurate because the HST fields probe nearly 
the entire luminosity function, because they require no uncertain sky
subtraction, and because they directly overlap several of the
spectroscopic fields.  We calculate the normalization using three
spectroscopic fields that overlap the HST fields, making a small
correction for the slight differences in mean radii between corresponding 
HST and spectroscopic fields.  We choose the normalization
value using a minimum $\chi^2$ criterion, with results shown on the plot.  
It can be seen that the total surface brightness in our model
traces observational data reasonably well, in particular reproducing
the previously known upward kink around 20--30~kpc \citep{irwin05,courteau11}.  
Although some discrepancies may
be present, these are no larger than the scatter in the observational
points.  Some of the observations are still likely contaminated with
substructure in the outer parts of the halo, so our model is chosen to
follow the lower envelope of the data.

Both bulge and halo components are represented here by spherical, 
isotropic populations.  We regard the distinction
between these hot bulge and halo components as somewhat artificial.
The evidence for two components in the surface brightness data is
merely the previously mentioned upward kink,
and the separation between bulge and halo in our model is
undoubtedly influenced by our choice of radial profiles.  Thus we
prefer to discuss them collectively as the ``spheroid'' component,
which is shown by the dashed red line in Fig.~\ref{fig.sb}.
The transition between the surface mass density of these
two components occurs at 25 kpc, and that in the $V$ light at 21 kpc.
We note the decomposition here is rather different than that for
the $I$-band decomposition in \citet{courteau11}, 
in which the bulge-halo transition occurs
at a mere 3~kpc, due largely to differences in the assumed radial profiles.
Courteau's preferred model uses a de~Vaucouleurs bulge and a cored
power-law halo, and these have very different dependences at small
and large radii than our functional forms.
The total spheroid components in our model and in the preferred model
of \citet{courteau11} are not so different, however.  Even with a crude
correction for $V-I$ color, we find the two profiles differ usually by
less than 0.3 magnitudes within the virial radius and never by as much
as a magnitude.  The two largest discrepancies come at the bulge-halo
transition points in the two models, and lie either in the region
where the disk dominates the spheroid, or in the halo region where the
statistical and systematic uncertainties are high.  

The spheroid star velocity dispersion at 30 kpc in our model
is $118 \kms$ and very
close to Gaussian.  From measurements elsewhere in M31's halo, we
expect a nearly Gaussian distribution with a mean of zero and
dispersion of about $130 \kms$ in the hot spheroid
\citep{chapman06,gilbert07}, suggesting the model spheroid component 
is reasonable.

We use equal particle masses for all components of our simulation,
so that the number density of simulation particles is proportional to
the mass density within the model.  In principle the ratio of mass
density to the number of RGB stars that are selected in the
observations is dependent on stellar population.  However, the bulge
and halo population at these radii have relatively old stellar
populations lacking in stars with $<5 \Gyr$
\citep{brown03,brown06b,brown07}, so their RGB density per unit mass
should not be too different from the stream material.  As we will see
later, the disk component plays at most a small role just in the 
innermost mask, so the possible large difference in its stellar
population is not of great concern.  Furthermore, accurate predictions
of the kinematic component normalizations are never required by our
analysis methods below.  Rather, we allow freedom in fitting the
normalizations of the different components, and instead use their 
spatial and velocity distributions as a template to set the observed
relative strength of the component contributions.

When comparing to the observations, we enhance our particle statistics
by using a larger area than is covered by the observed masks.  We
choose particles within the region with major axis position $|X| <
0.3\degree$, slightly wider than any of our masks, as well as minor axis
position $0.7\degree < Y < 2.2\degree$.  In this we rely on the fact
that over small scales in the simulation the velocity distributions
change slowly with position, and depend mainly on projected radius.

\section{ANALYSIS AND RESULTS} 
\label{sec.analysis}
\subsection{Observed distributions}

\begin{figure}
\includegraphics[width=8.4cm]{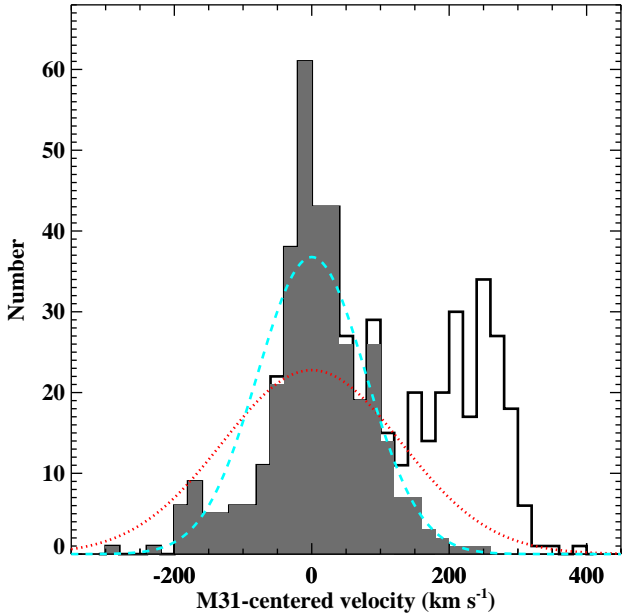}   
\caption{
\label{fig.vhist}
Histogram of the stellar velocities in our sample.  The solid
histogram shows only those stars classified as M31 giant stars, while
the open histogram includes the contribution from foreground dwarfs.
The dotted line shows the best Gaussian fit to the M31 stars assuming 
the expected spheroid velocity dispersion of $130 \kms$, while the dashed
line shows the best fit with velocity dispersion left free.  Neither
distribution adequately fits the data.}
\end{figure}

\begin{figure}
\includegraphics[width=8.5cm]{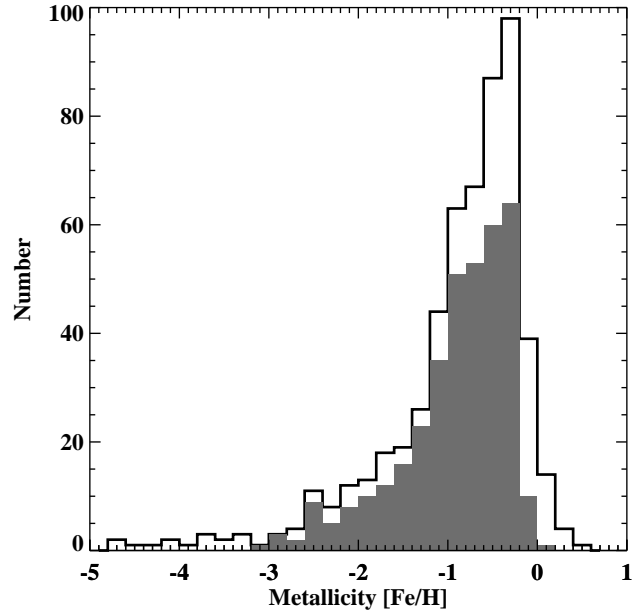}
\caption{
\label{fig.methist}
Histogram of the stellar metallicities in our sample, with the
solid histogram showing only the M31 giant stars.
We use photometric metallicities throughout, since we believe 
these to be more reliable than spectroscopic metallicities at
the noise level of our spectra.  
}
\end{figure}

\begin{figure}
\includegraphics[width=8.5cm]{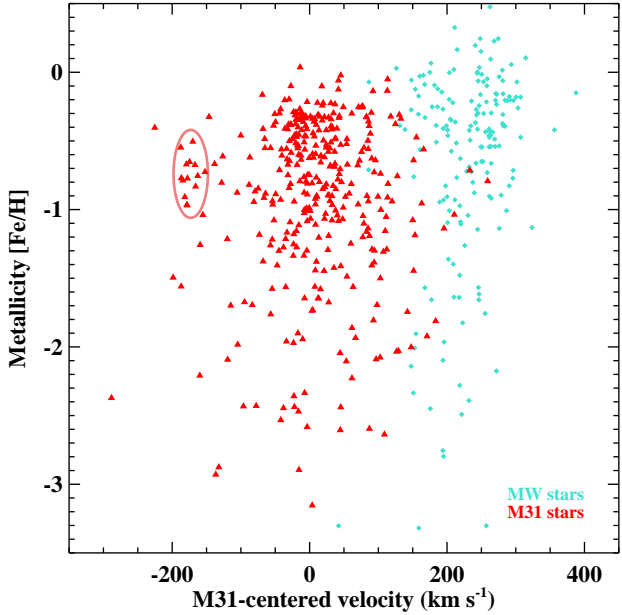}
\caption{
\label{fig.vmet}
Metallicity versus velocity for individual stars in our sample.
Triangles denote M31 giant stars, and diamonds denote MW dwarf stars.  
(Assigned metallicities of the MW stars are not meaningful
as explained in the text.)
We have placed an ellipse around the
small clump at $v \approx -175 \kms$,  $\mbox{[Fe/H]} \approx -0.75$.  
}
\end{figure}

\begin{figure}
\includegraphics[width=8.5cm]{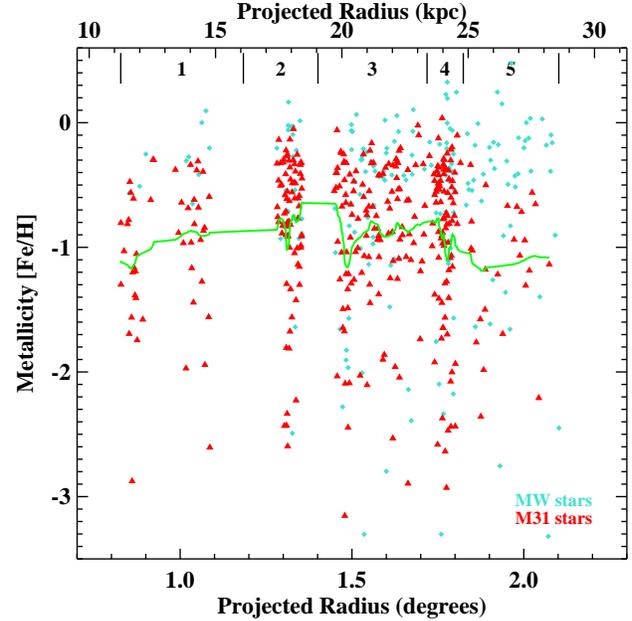}
\caption{
\label{fig.rmet}
Metallicity versus radius for individual stars in our sample,
with symbols the same as Fig.~\ref{fig.vmet}.
The green line shows the metallicity boxcar smoothed over 
radial rank order.  The numbers and separators
near the top indicate the radial zones.
}
\end{figure}

\begin{figure}
\includegraphics[width=8.5cm]{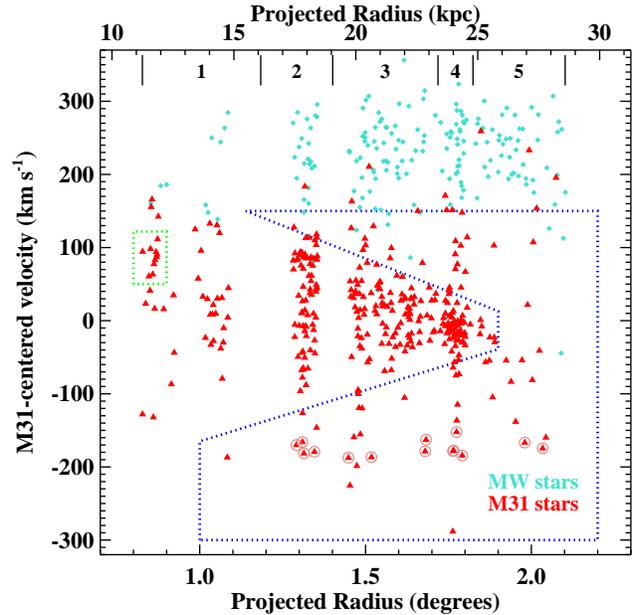}
\caption{
\label{fig.rvobs}
Velocity versus radius for individual stars in our sample.  Red 
triangles denote M31 giant stars, and cyan diamonds denote MW dwarf stars.  
The stars within the ellipse in Fig.~\ref{fig.vmet} are circled.
The green box indicates a clump of stars possibly associated with NGC 205.
The blue polygon shows the test region used to calculate the 
velocity dispersion of the hot component.
}
\end{figure}

In Fig.~\ref{fig.vhist}, we show the overall radial velocity distribution 
of the observed stars.
Throughout, we state velocity relative to M31's systemic
velocity of $-300 \pm 4 \kms$ \citep{devauc91}.
We split the distribution into M31 and MW stars.
As discussed in \S~2, 
we expect a mean of zero and dispersion of $\tsim 130 \kms$ if 
the distribution is dominated by the hot spheroid.
Neither this distribution, nor the best Gaussian fit 
with dispersion left free, fits the distribution at all well, 
as can be seen in the figure and as verified by $\chi^2$ tests.
Most stars are contained within $|v| < 50 \kms$, but
a large fraction inhabit much broader tails of the distribution.
In contrast, a fit with dispersion $130 \kms$ to only the values with 
$|v| > 100 \kms$ is statistically acceptable.
The apparent peaks near $-175 \kms$ and $-20 \kms$ hint at
finer-grained structure.

In Fig.~\ref{fig.methist}, we show the distribution of our
sample stars in metallicity.  The stars are predominantly
metal-rich, though a tail to low values is evident.  
In Fig.~\ref{fig.vmet}, we show the metallicity versus velocity
of individual stars,
color-coded by classification into M31 vs MW stars.  
There is a small clump at $-175 \kms$ which is the main contributor to 
the leftmost hump in Fig.~\ref{fig.vhist}.  
Note also that at velocities close to M31 systemic, there 
is a strong concentration of high-metallicity stars.
This results in metallicity distributions of the stars with $|v|<75 \kms$ 
and those with $75 \leq |v| <150 \kms$ being inconsistent at the $<3\%$ 
level, according to a Kolmogorov-Smirnov test.

In Fig.~\ref{fig.rmet}, we show the metallicities of individual
stars as a function of projected radius, with the median trend
overlaid.  The coverage gaps and variations in radial density are due
to the positioning and orientation of the masks: the two tangentially
aligned masks produce narrow concentration of stars. The four radially
aligned masks (with two overlapping) produce three broader but less
dense zones of stars.  The green line shows the giant-star
metallicities boxcar smoothed over their rank order in radius.  This
seems to indicate a rapid change in the metallicities around the edge
of the shelf at about $1.83\degree$, indicating a change in the mean
population there.  Comparing the metallicities of M31 stars in zones
4 and 5, we find a 97\% chance using bootstrap resampling that the 
mean is higher in zone 4, which supports the visual impression.

In Fig.~\ref{fig.rvobs}, we show the velocity versus radius of 
the sample stars.  The uneven distribution in radius reflects 
the effect of the mask pattern.
The velocity distribution of the M31 stars varies with radius.
For example, a Kolmogorov-Smirnov test indicates that the distributions
within radial zones 2 and 4 are inconsistent at the $<10^{-4}$ level.  
The velocity distribution 
is very far from Gaussian at most radii.  Clumps of stars are
obvious at about $R=1.33\degree$, $v=75 \kms$, and $R=1.77\degree$,
$v=-12 \kms$.  There also appears to be a wedge-shaped region within
which many of the other stars reside, with these two clumps at two of
its vertices.  These velocity clumps are not especially concentrated
in position within their masks, and are clearly not the result of an
individual star cluster.  
The large number of M31 stars outside
the wedge-shaped region also clearly shows a smoother hot spheroid is
present.  

A small overdensity of 10 stars around $R=0.85\degree$, $v=89 \kms$ is
indicated by a green box.  Observations by \citet{geha06} indicate NGC
205 has a tidal debris tail extending roughly in the direction of our
innermost mask, and at their furthest detected extent 
of about 3.6~kpc from NGC 205's center the M31-centric
velocity of this debris is about $80 \kms$.  
The stars in the green box span a range of 3.2--3.8~kpc from
NGC~205's center.  The mean $\feh$ of stars
in the green box is $-1.2$.  For comparison, the old stellar
population in NGC 205 has a mean of $-1.06 \pm 0.04$ according to
\citet{butler05}, consistent with our stars within the uncertainties.
Thus the kinematics and metallicity strongly suggest this overdensity
is a continuation of a tidal tail from NGC 205.
It remains to be seen whether the debris tail found here and in
\citet{geha06} can be reconciled with the suggested kinematic 
detection of a tail extending from NGC~205 to the NE \citep{alan04}.

This overdensity helps to explain the asymmetric velocity
distribution in our innermost mask, where the stars are mainly at
positive velocities.  We note the innermost stars are at about
$0.8\degree = 48 \kpc$ in the disk plane.  It is possible that the
disk contributes to some of the low-velocity stars in this mask,
though a warp might be necessary to explain the offset from zero
velocity.

The strong wedge-shaped overdensity of stars is more significant and
more challenging to explain in terms of known kinematic components of
M31.  Since the observations are taken along the minor axis, one 
would expect any contribution from M31's disk to have nearly zero velocity.
Even the extended cold disk component found by \citet{ibata05}, 
which has a substantial lag in rotation speed relative to the cold gas 
in the disk, cannot produce a $75 \kms$ line-of-sight velocity relative 
to M31, as exhibited at $1.33\degree$.  The outer cold clump is at an
appropriate velocity to be from the disk, but this location is at a
projected radius of $24 \kpc$ along the minor axis, or $107 \kpc$ in
M31's disk plane (for an inclination of $77\degree$).  
There is no sign of a disk contribution in the spectroscopic masks 
interior to the outer cold clump, or in the photometric survey maps
(Fig.~\ref{fig.masks}), so we conclude the outer cold clump is not
associated with the disk either.

As mentioned above, the disk contribution in the range 10--20~kpc on
the SE minor axis is controversial.  Kinematic observations suggest
dominance by the spheroid \citep{kalirai06b,gilbert07}, while
photometric observations have been interpreted as dominance by a disk
component \citep{ibata07}.  However, there seems to be agreement that the 
disk contribution is negligible beyond 20 kpc.  Our kinematic
results show this is also true from 18--24 kpc on the NW minor axis.

To test the velocity dispersion of the hot component outside the
triangular enhancement, we create a test region as shown in
Fig.~\ref{fig.rvobs}.  We carry out a simple Bayesian
analysis of the velocity dispersion of the stars in this region,
assuming they have a hot Gaussian velocity distribution centered at
zero and with width $\sigma_v$, and taking a flat prior for
$\ln(\sigma_v)$.  This yields an estimate of $\sigma_v = 117 \pm 15
\kms$, with the maximum likelihood value occurring at $113 \kms$.
Using a more conservative test region makes no significant change
in these values.  The hot component dispersion found here is consistent
with the dispersion $\sigma_v \approx 130 \kms$ expected from previous
studies in other regions around M31 at slightly larger radii
\citep{chapman06,gilbert07}.  It is also in excellent agreement
with the spheroidal component of M31 in our model, which has a
dispersion ranging from 103 to $118 \kms$ over the range 20--$30 \kpc$.

\subsection{Comparison with model kinematics}
\begin{figure}

\includegraphics[bb=14 32 481 556,width=8.5cm]{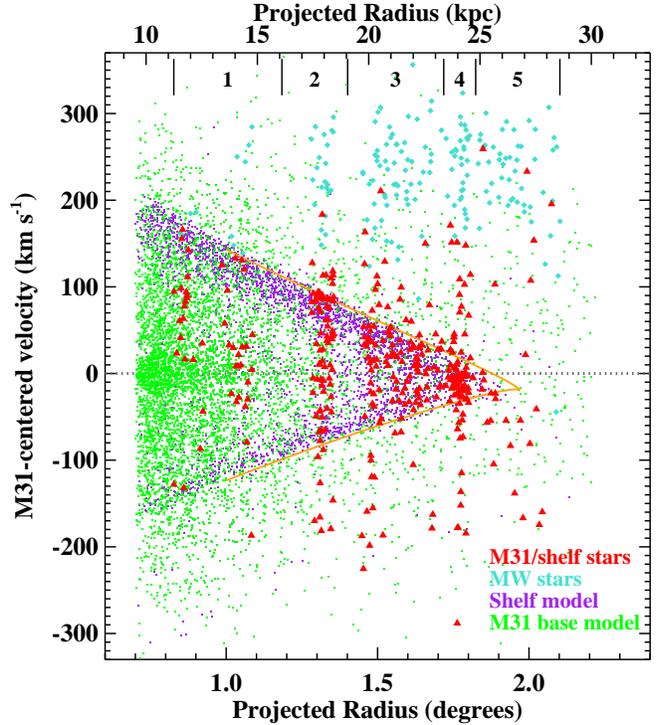}
\caption{
\label{fig.rvcomp}
Velocity versus radius as in Fig.~\ref{fig.rvobs}, but this time 
including our simulation particles as small points.
Particles are selected from a rectangle around
the masks as discussed in \S~\ref{sec.sim}.
Purple points show the shelf component formed by the 
GSS satellite debris,
while green points show the component representing M31 itself,
which is dominated by the hot smooth spheroid component.
Red observed stars are to be compared to the combination of 
green and purple model points.
The orange lines show the analytic envelope discussed
in the text.  
}
\end{figure}

\begin{figure*}
\includegraphics[bb=17 7 495 420, width=6.004cm]{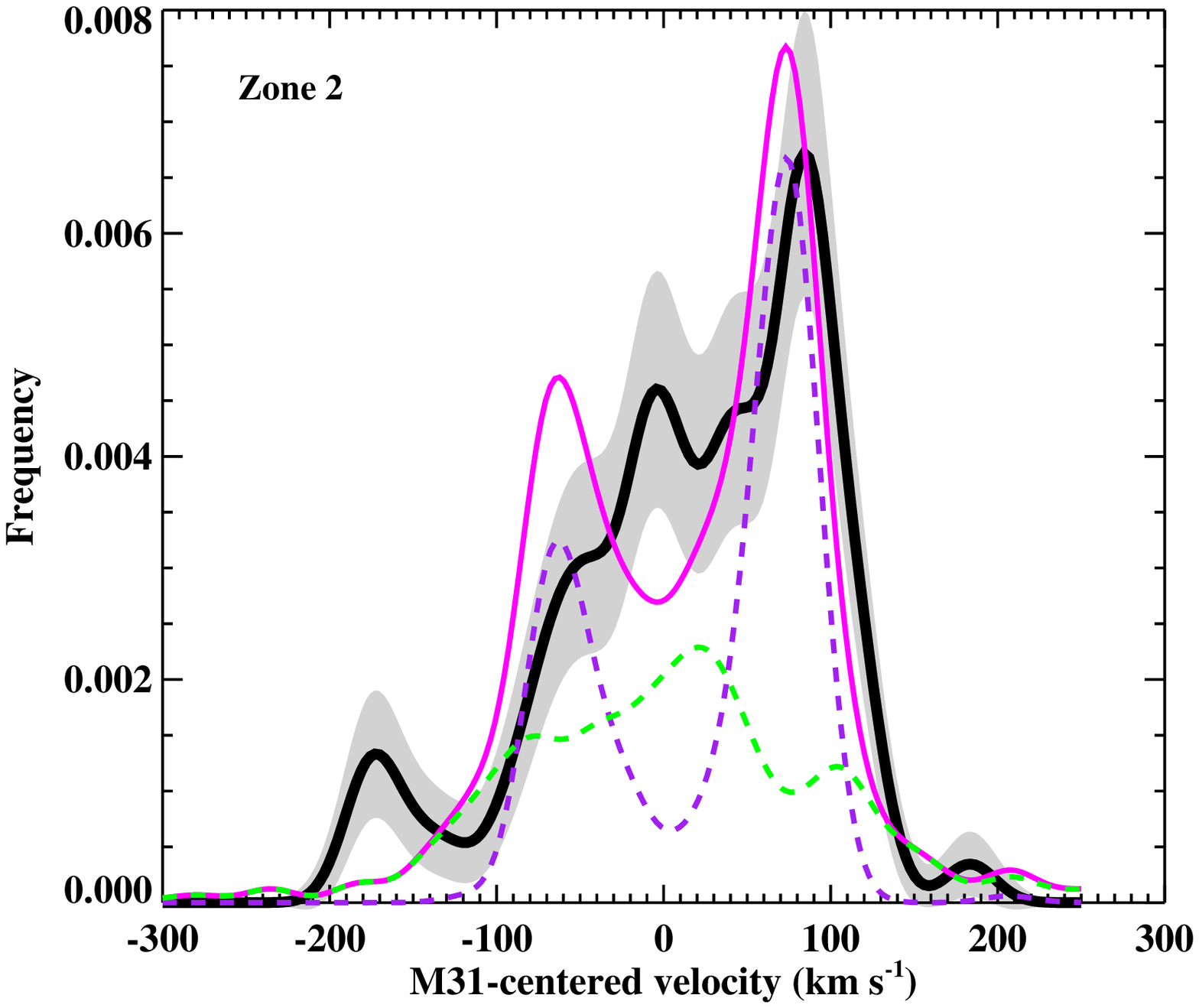}  
\includegraphics*[bb=36 7 495 420, width=5.765cm]{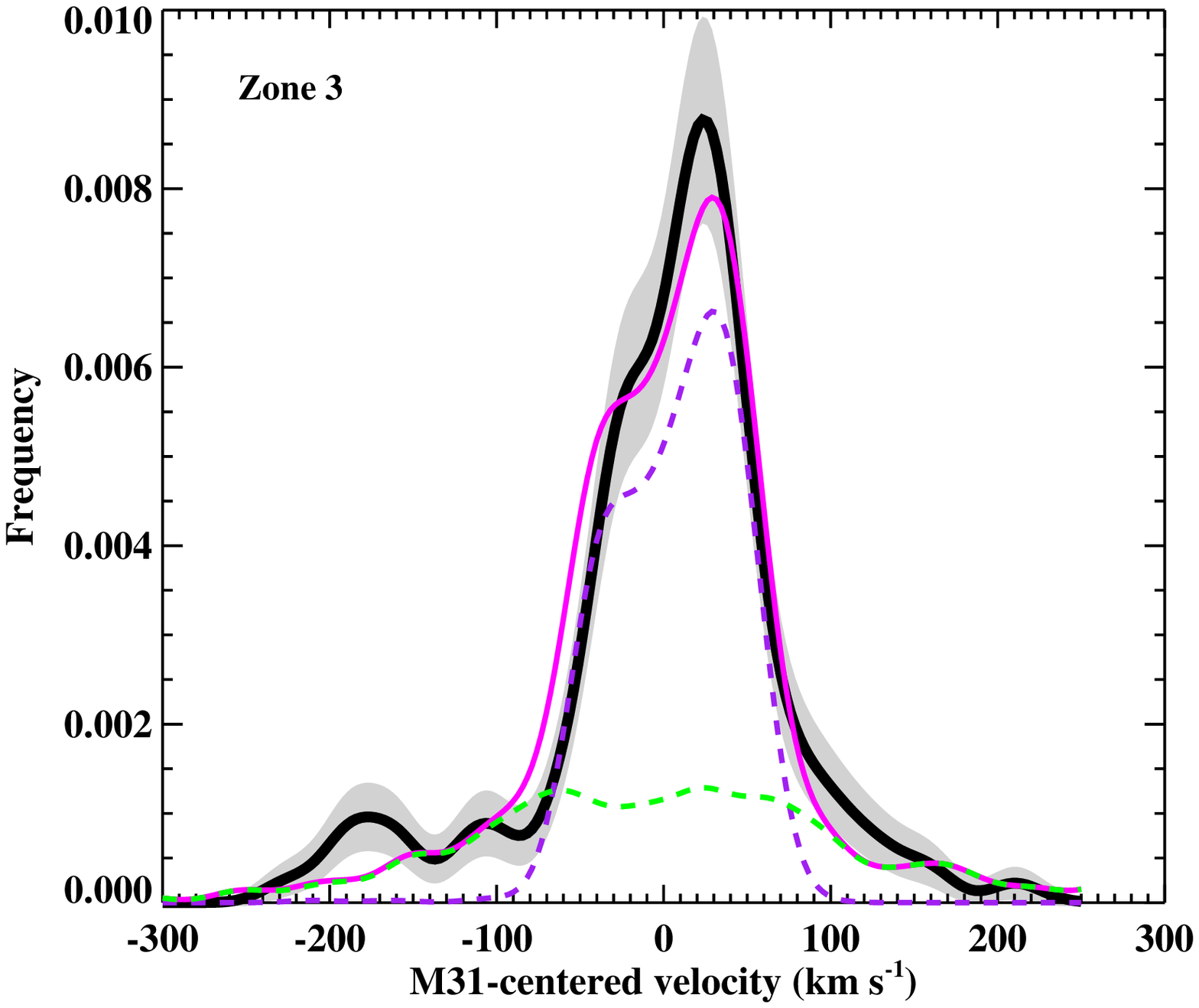}
\includegraphics*[bb=36 7 495 420, width=5.765cm]{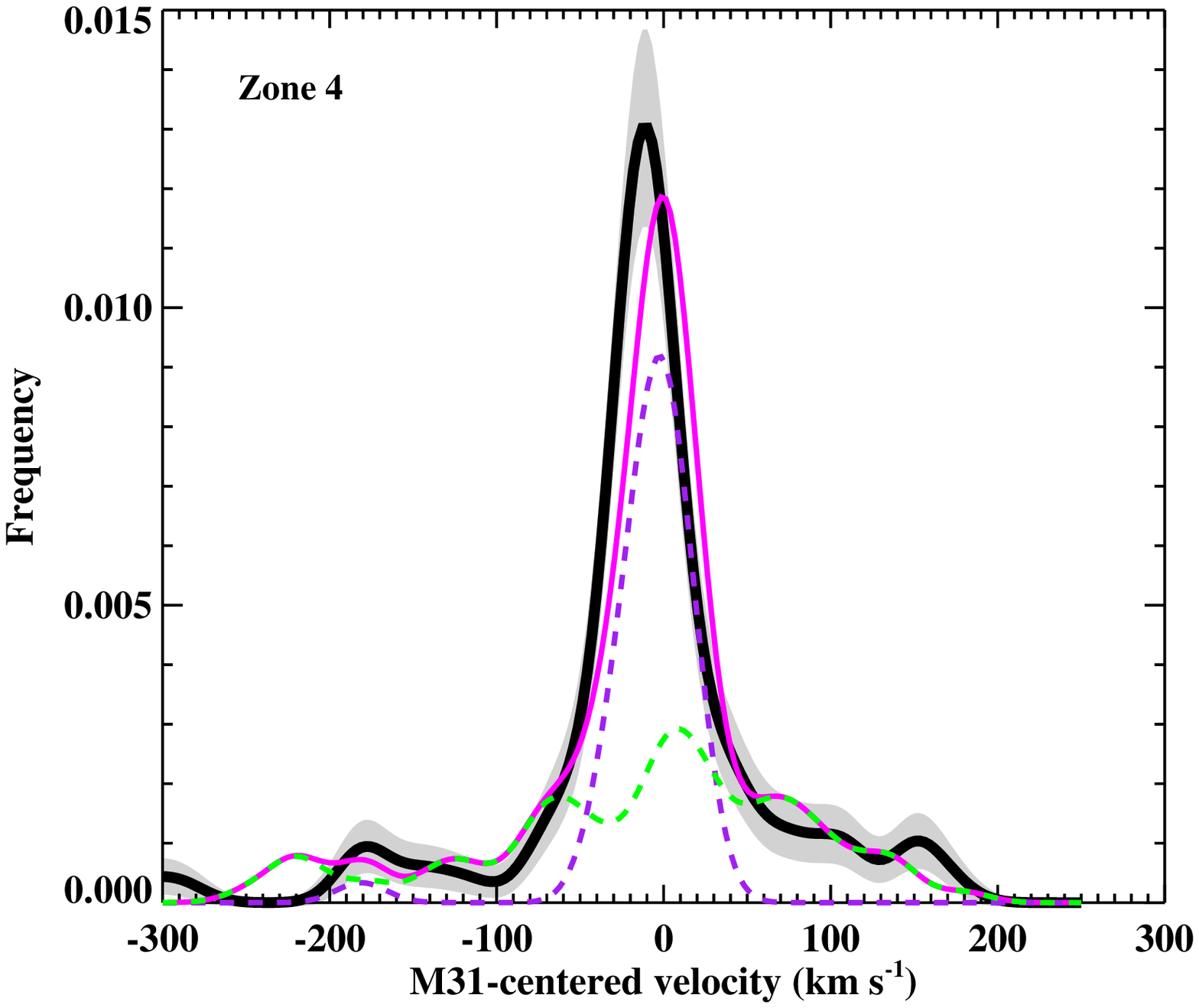}
\caption{
\label{fig.vdist}
Distribution of M31 stars in velocity
within radial zones 2 (left), 3 (center), and 4 (right).
Observations and model are kernel-smoothed with a Gaussian
of $\sigma=15 \kms$.  Black line: observed M31 stars.
The shaded region indicates the $1\sigma$ error band.
Magenta line: total distribution in model.  
Purple line: GSS-related model stars.  Green line: M31 galaxy model stars.
}
\end{figure*}

In Fig.~\ref{fig.rvcomp}, we again show the velocity versus radius
of our stars, this time overlaid on the simulation points from the 
surrounding area.  
Purple points show the particles belonging to the GSS-related debris, 
and green the particles from the base M31 model.  
For slightly better agreement we have taken the liberty of 
reducing the radii of the GSS debris particles
by 4\%, an amount well within the model uncertainties 
(Fardal \ea in preparation).  The agreement between
the simulation prediction and observations is striking.  
The simulated GSS debris lies in a wedge that tapers to a tip at about
$1.9\degree$ or 26 kpc, with a strong concentration at the upper edge.
The wedge shape results from a caustic in projected velocity that 
is a consequence of the kinematics of a radial shell
\citep{merrifield98}. A large fraction of the observed stars are
distributed in a similar pattern.

In the rest of this subsection we will compare the distribution in the
radius-velocity plane in more detail.  Rather than just making a statistical
comparison in arbitrary areas of this plane, we will try to interpret
the observed results in a physical way.  
We first note that the slope of the observed wedge appears to agree with 
the simulation.  Since this slope is related to the gravitational force
at the shelf edge \citep{merrifield98}, this suggests our gravitational
potential model near radius $20 \kpc$ is reasonable.

The wedge shape of the caustic derived in \citet{merrifield98} is triangular,
and symmetric around zero systemic velocity.  However, this result
assumes that the shell star orbits are exactly radial (no angular
momentum) and monoenergetic (no velocity dispersion or energy gradient
along the flow).  In contrast, our simulated shelf material has both
angular momentum and an energy gradient along the flow.  In the
Appendix we derive an analytic expression for the caustic envelope that
incorporates both of these effects, as well as second-order terms in
the potential and velocity projection.   To use this expression we need 
parameters for the rotation and energy gradient which we now consider.

The mean observed velocities are about $-12 \kms$ in the large group
of cold stars in zone 4 just inside the shelf tip, and $-18
\kms$ for the smaller group of six stars at slightly larger radius
and velocity near zero.  Our simulation of the debris similarly
assigns small negative velocities for these regions.  We estimate the
tip velocity in our simulation as about
$v_{tip} \approx -20 \kms$, by taking a sample of particles at the
shelf edge.  In contrast, a shell model assuming purely radial motions
for the stars would find a zero velocity relative to the host, at
least to the measurement precision of the systemic velocity which here
is roughly $4 \kms$.  The small negative velocity we find in the
simulations has clear significance: it represents the effect of
angular momentum within the shell particles, which causes motion along
the line of sight at the shell edge.  We infer the same is true for
the observed stars.  The direction of the angular momentum results
from the specific trajectory assigned to the GSS progenitor in the F07
model, in which stars after passing through the NE Shelf pass behind
M31 into the W Shelf, curling toward us at the edge of the shelf.
This trajectory is strongly constrained by the need to place the
stream and shelves correctly on the sky.

In the simulation, the radial shell giving rise to the W Shelf can be
seen to expand over time as stars of higher and higher orbital
energies arrive there.
This energy gradient is expected on quite general
grounds, since stars with higher orbital energies take longer to
complete their orbits.  
The energy gradient skews the
wedge shape, increasing the apparent speed of outflowing stars
(moving them further from M31's velocity)
and decreasing that of the inflowing stars.
The simulated shelf stars mostly lie beyond the midplane, 
so that outflowing stars have positive velocities
and inflowing ones have negative velocities; 
this results in an upward skew of the wedge.

To combine the skewed velocity envelope given by equation
\ref{eqn.causticapprox} with a global offset due to the angular
momentum, we need four parameters.
We estimate the shelf expansion speed $\vsh = 40 \kms$
from the evolution of the shelf radius in the simulation.
We use $v_{rot} = -20 \kms$ as the result of the line-of-sight
projection of the angular momentum, as found above.  
We choose $\rsh = 1.97\degree$ to match the observed 
three-dimensional shelf radius.
Finally, from our potential model, we estimate the
circular velocity $V_c = [G M(r) / r]^{1/2}$ 
to be $240 \kms$ at the shell radius.  
These choices yield the orange lines shown in Fig.~\ref{fig.rvcomp}.
Clearly this envelope is a good match to the actual envelope of both
the simulation and observed stars.  In contrast, a symmetric linear
wedge would be a poor fit to the simulated and observed caustic shapes.  
The upward skew of the wedge towards M31's center suggests that the
simulation correctly places the bulk of the shelf beyond the midplane,
so that the positive-velocity stars are the outflowing ones.

\begin{figure}
\includegraphics[bb=15 9 496 425,width=8.5cm]{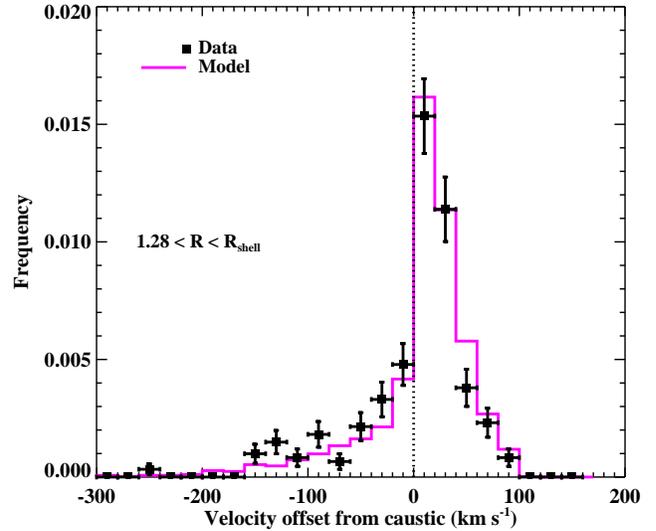} 
\caption{
\label{fig.vhistcomp}
Binned distribution of velocity difference $\Delta v$ from the caustic
envelope shown in Fig.~\ref{fig.rvcomp}, as defined in the text.
Points with error bars show the observed distribution.  The magenta
line shows the simulated distribution.  The significant jump at zero
demonstrates the increase in density just inside the velocity caustic.
}
\end{figure}

In our simulation the M31 disk contributes a small amount to only the
innermost mask, but in actuality it is not obvious that any stars with
radii and velocities typical of the disk are present in our sample.
Therefore our comparison to observations constrains only the shelf
and spheroid components.
We regard the particular bulge/halo decomposition of
the spheroid in our model as somewhat arbitrary, 
Formally, though, our M31 bulge component dominates the contribution 
to the surface mass density of the spheroid component 
out to $25 \kpc$ or $1.8\degree$, in other words nearly
the entire region probed by this survey, while the M31 halo
component dominates outside that radius.

In Fig.~\ref{fig.vdist} and Fig.~\ref{fig.vhistcomp}, we compare
in greater detail the distributions of 
simulated and observed stars in velocity space.
We ignore zone 1 in all further comparisons with the simulation, 
due to the small number of detected stars and the apparent contamination 
from NGC 205 debris and possibly M31's disk.  
(See Fig.~\ref{fig.masks} and Table~\ref{table.masks}
for the positions of the radial zones.)

Fig.~\ref{fig.vdist} shows smoothed-kernel representations of
observed and simulated stars in the three central zones.
In zone 2, there is clearly a steep-sided central component in 
both observations and simulations, weighted strongly to
positive velocities.  The simulation is more strongly peaked,
and the smaller negative-velocity peak is not clearly distinct
in the observations.  This could be from an imbalance between our
spheroid and shelf components, or a slightly different spatial
distribution of shelf stars in this range.
In zone 3, the similarity between the simulation and the observations
is remarkable.  As we approach the edge of the shelf in zone 4, 
the central component shrinks to a cold peak.  
The stars in the observed and simulated cold components actually occupy
a similar range, but a slightly different distribution of stars
within this range produces a small shift in the peak of the
smoothed velocity distribution.
Clearly, the model provides an essentially correct description of 
the motions of stars within the shelf.  

The sloping caustics shown in Fig.~\ref{fig.rvcomp} are smeared over
velocity in Fig.~\ref{fig.vdist} by the finite range in radius
within each zone.  Fig.~\ref{fig.vhistcomp} instead shows the
velocity distribution in reference to this caustic velocity.  We use a
histogram rather than kernel smoothing to make a sharp division
between stars inside and outside of the caustic.  For each observed or
simulated star $i$ with $1.28 \degree < Y_i < 1.9\degree$ (i.e.,
from the inner edge of zone 2 to the shelf boundary), we compute a 
quantity
$\Delta v_i = \pm (v_{env,i}-v_i)$, 
where $v_{env,i}$ is the nearer of the upper and
lower caustics given by the orange lines in
Fig.~\ref{fig.rvcomp} at the star's radius.  We use the plus (minus)
sign for stars nearer the upper (lower) caustic. Thus $\Delta v_i$ is
small and positive for stars just inside the caustic, and small and
negative for stars just outside.  We show the observed distribution as
points with Poisson error bars, and also the result of the simulation
as a solid histogram.  It is clear that the density jumps by a large
amount as one crosses the caustic boundary.  The simulated
distribution reproduces this distribution rather well, although
the fit is poor from a statistical point of view.  The same can
be said of the $\Delta v$ distributions in individual fields which 
are not shown here.  Not surprisingly, the shelf component is
solely responsible for the jump at zero in the simulation, while the
M31 component has a broad, smooth distribution.

Now that we have clearly established the presence of both hot smooth
and shelf components from their velocity distributions, we use these
distributions as templates to find the observed relative strengths of
these component.  For each of our radial zones, we find the GSS and
M31 velocity distributions at the velocities of the observed stars by
kernel-smoothing the simulation velocities.  We combine these two
distributions in proportion to their total masses within the zone,
with the GSS component weighted by an additional factor $g_{sat}$.  
We then compute the total likelihood $\ln L_{sat}$
for the M31 stars being drawn from this distribution, by adding 
the logarithms of individual velocity distribution values.
Repeating this process for a set of values for the weighting
factor $g_{sat}$, 
we find its maximum likelihood estimates and error estimates,
and we can compare the likelihood $\ln L_{sat}$
to that of the null model $\ln L_0$, where we use only the
M31 component and no GSS debris (equivalent to setting $g_{sat} = 0$).
This test offers very strong evidence for a satellite component in 
zones 2--4, whereas zones 1 and 5 are ambiguous. 
(This is not surprising since NGC 205 contaminates the innermost
masks, while the outer mask should have very few shelf stars.)
In our first set of tests we let $g_{sat}$ take on a different
value for each zone.  For zones 2, 3, and 4 respectively, 
we find optimum satellite weights $\ln g_{sat}$ of 
$-0.8 \pm 0.5$, $-0.1 \pm 0.3$, and $0.5 \pm 0.3$.
The log likelihoods relative to the null model are
$\ln(L_{sat}/L_0)$ of 8, 29, and 31, indicating
solid detections of the GSS component in each of these zones.
If we assume a constant $g_{sat}$ and combine data from all fields,
we find $\ln g_{sat} = 0.0 \pm 0.2$, i.e.\
the optimum weight of the GSS component is 
$\bar{g}_{sat} = (1.0 \pm 0.2)$ the standard weight in our 
model. \footnote{One should not read too much into the fact $\bar{g}_{sat}$
is near 1 here, as we used the considerable observational freedom 
in the spheroid modeling of \S~2 to choose a model where this 
was assured.  If we normalize the GSS component by this weighting factor,
however, all of our other results are virtually independent of which spheroid 
model we use.}
We note that $\ln(L_{sat}/L_0) = 72$, i.e. 
in a simple likelihood ratio interpretation 
the kinematic distribution is $>10^{31}$ times more likely 
to be drawn from the model with a satellite component than that without.
This overwhelming statistical requirement for a shelf kinematic component 
confirms the visual impressions from
Figures~\ref{fig.rvcomp} and \ref{fig.vhistcomp}.

For future work on modeling the GSS, it will be useful to have an
estimate of the contribution of the shelf to the M31 stars.  
The purely photometric sample of \citet{tanaka10} shows a steep
surface brightness drop at the W Shelf edge, which we can estimate
to be roughly 3 magnitudes in their most metal-rich channel
by extrapolating their outer surface brightness profile inwards.
This would suggest the spheroid constitutes approximately 0.06
of the total metal-rich stars and the shelf component itself 
about 0.94, albeit with large uncertainties.

The ratio of shelf to spheroid components varies with position, 
and the average ratio obtained with our somewhat irregular spatial 
coverage is of no particular interest.
Therefore we use the simulations to estimate this ratio averaged 
over an area defined by the 
radial range $1.35$--$1.95\degree$ and the azumuthal range 
0--$30\degree$ to the SE of the NW minor axis.
We compute the satellite-contributed fraction $f_{sat}$ 
from the numbers of simulation particles in this region 
in the GSS and M31 components, weighting the GSS count
by the satellite weight $\bar{g}_{sat}$.  Using the likelihood
as a function of $\bar{g}_{sat}$ just described, we find the GSS
stars make up $f_{sat} = 0.59 \pm 0.05$ of the total stars
in this region.
Better definition of the GSS-related structures can be obtained
by focusing on the metal-rich population.
If we repeat the likelihood calculation using only metal-rich
stars with $\feh > -0.75$ to constrain $g_{sat}$,
we require a higher weight of $\bar{g}_{sat} = 1.34 \pm 0.37$.  
This yields a higher satellite fraction of $f_{sat} = 0.84 \pm 0.05$
for this subset of stars.
(This suggests that the shelf is more metal-rich than the
spheroid, which we discuss in more detail later.)
These estimates are the most precise determinations yet of the ratio 
between the shelf and spheroid, and should be very useful in constraining
future models of the GSS.

The observed velocities appear somewhat asymmetric about the
origin; most noticeably, the strong enhancement at 
$R=1.33\degree$, $v=75 \kms$ is not matched by an equally
strong clump at $-75 \kms$.  The model shelf component
shows an asymmetry in the same direction (compare the
strengths of the upper and lower caustics in 
Figures~\ref{fig.rvcomp} and \ref{fig.vhistcomp}).  
The simulated shelf stars lie beyond the midplane,
and we have argued above the observed stars must as well
due to the skew of the wedge in $r$-$v$ space.
Therefore the positive velocity stars are outgoing and
the negative velocity stars are infalling.  
The ratio of these two components tests the density
gradient along the stream, since the infalling stars
have larger orbital phase.

To test the model versus the data in a more precise way we restrict
our attention to zones 2 and 3, which have a large fraction of shelf
stars but are far enough away from the edge that the precise dividing
line between outgoing and infalling stars is not important.  In the
simulation, the outgoing stars (classed as such by their actual 3-d
radial velocities) make up $(67 \pm 1)$\% of the shelf stars within
this region.  The dividing line between outgoing and infalling stars
is reached here at $-8 \kms$.  We use the GSS and M31 velocity
distributions, along with the maximum likelihood satellite debris
weight factor of $f_{sw} = 0.71$, to assign a GSS mixing fraction to
each observed star.  We can then estimate the number of outgoing and
infalling GSS stars by summing up these mixing fractions above and 
below the dividing line of $-8 \kms$.  
The observed outgoing fraction is $70 \pm 3$\%,
in excellent agreement with the simulation.  The density therefore 
falls off steeply with increasing orbital phase.  If we accept that
the stream density should rise when approaching the phase of the
progenitor, then this demonstrates the shelf lies at a larger orbital 
phase than the main body of the progenitor, as in our model.

\subsection{Metallicity of shelf and spheroid components}
Our maximum-likelihood weighting of GSS and M31 components over all
zones produces for each observed star a mixing fraction, that
is, a probability that the star is from one component or the other.
We now use these mixing fractions to estimate the metallicity distributions
of the individual components.  

\begin{figure}
\includegraphics[bb=27 8 490 420,width=8.5cm]{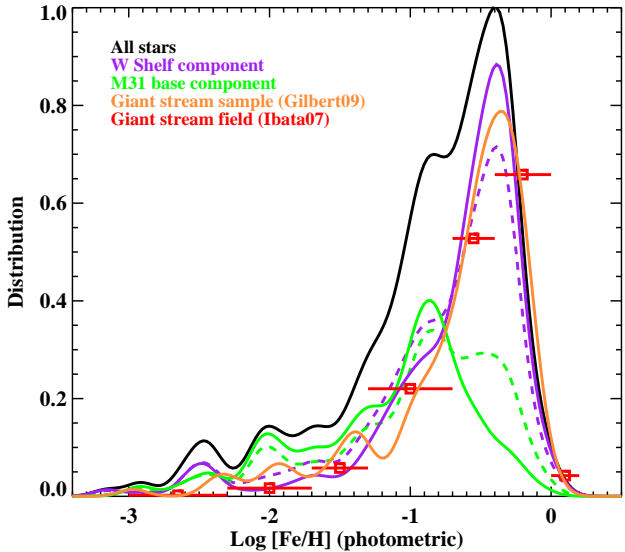}
\caption{
\label{fig.metmix}
Kernel-smoothed distribution of our W Shelf sample stars
in metallicity (solid black curve).  The dashed curves indicate the
contribution of the shelf (purple) and M31 proper (green) components,
using the initial decomposition based on kinematics.  The solid curves
with the same colors indicate the contributions using the mixing
fractions based on both kinematics and metallicity.  The observed metallicity
distribution of stars within the GSS itself, about 100 kpc to 
the south, is shown by the orange kernel-smoothed curve
for \citet{gilbert09}, and by the squares with error bars for \citet{ibata07}.}
\end{figure}

In Fig.~\ref{fig.metmix} we show the metallicity distribution
for the entire sample of M31 stars, kernel-smoothed with
a Gaussian of dispersion 0.1 dex.  Dashed colored lines show the
contribution of the individual components, 
where each star contributes an amount proportional to its
mixing fraction for a given component and is kernel-smoothed as before.
It is clear that the GSS
debris has higher metallicity on average than the smooth spheroid.

Our kinematic separation of the stellar components is far from
complete---there are many stars with significant mixing fractions in both
components, meaning the metallicity distributions of the components
are still to some degree mixed together.  This mixing can be reversed
by applying the metallicity distribution as an additional criterion in
computing the mixing fractions.  Here we assume that metallicity is
independent of position and velocity throughout our observed
region.\footnote{A constant metallicity distribution in the observed
phase space is probably a good assumption for the W Shelf material
itself, since the stars are moving rapidly through the radial shell.
For the M31 model, the range of radii we sample is less than a factor
of 2 and the spheroid dominates over the entire range, so the
distribution in this component is probably reasonably constant as well.}
Starting with the original mixing fractions from kinematics alone, we
compute the metallicity distributions of each component and use it as
an additional factor in the likelihood of a star being from that
component.  We then recompute kernel-smoothed metallicity distributions from 
the new mixing fractions, and iterate this process until the mixing fractions
converge.  In technical jargon, this procedure involves a non-parametric 
mixture model with prior fuzzy classifications on all the sample points,
which we solve for the mixture components using an 
expectation-maximization algorithm. The result is shown as the solid 
purple and green curves
in Fig.~\ref{fig.metmix}. This algorithm always amplifies the
differences between the original component distributions.  
It naturally amplifies noise as well, so we constructed mock samples from 
our simulation of similar size to the observational sample and tested the
separation using the remaining simulation particles to define the
components.  We found that at our sample size, the algorithm does a
good job of recovering the original component distributions without
introducing much visible noise.

Our final metallicity distributions in Fig.~\ref{fig.metmix}
show an even clearer separation of the satellite debris and smooth
spheroid components.  The mean metallicities are $-0.7$ for the satellite
debris and $-1.2$ for the spheroid component, for a typical difference of
$0.5$ dex in $\feh$.  
The spheroid metallicity has a similar mean and
dispersion to that found in the smooth, dynamically hot spheroid population
at slightly larger radii (30--$33 \kpc$) on the opposite side of M31 by 
\citet{kalirai06b}.  

The figure also illustrates the distribution of metallicity within the
GSS itself, from the kinematic sample of \citet{gilbert07} and the
ground-based photometry of \citet{ibata07}.  The \citet{gilbert07}
sample has advantages of kinematic selection and identical reduction
procedures to those employed in our W Shelf sample, but in truth the
two GSS distributions are quite similar.  The W Shelf debris component
shows remarkable agreement with these GSS samples, obtained in a
region $\tsim 100 \kpc$ away.

\section{DISCUSSION} 
\label{sec.discussion}
The observations presented here reveal two main components besides the
foreground contaminants from the MW: a hot smooth spheroid with a
roughly Gaussian velocity profile, and a shelf component with
kinematics matching that of a radial shell.  
The kinematic properties of the shelf component match
very well to a simulation of the GSS's formation. The metallicity
distributions of the two components are distinct, with the shelf
having higher mean metallicity and matching very well to observations
of the GSS.  
This agrees with the earlier conclusion of \citet{tanaka10},
who compared purely photometric samples in the W~Shelf and GSS
regions and also found them to be nearly identical.

Some caution is warranted when drawing connections between tidal
structures based on their stellar populations.  Different tidal
progenitors can have similar metallicity distributions, and a single
tidal structure can have gradients along it, as seems to be the case
for the Sagittarius stream \citep{chou07}.  Indeed, the GSS itself
seems to have a metallicity gradient from its core to its less dense
``cocoon'' on the SE side \citep{ibata07}.  The models of
\citet{fardal08} reproduce such a gradient by assuming a gradient in
the progenitor.  Importantly, however, neither observations nor models
suggest a strong gradient {\it along} the stream, especially for
regions separated by a comparable amount from the progenitor's core.
In the F07 model, the W Shelf leads and the GSS trails the current
core location by similar amounts.  Thus it is reasonable to connect
the two structures by their stellar populations.  

The precision of resolved-star spectroscopy allows a remarkable range
of conclusions to be drawn from the velocity structure.  We showed the
observations agree not just with the overall wedge pattern of the shelf
component, but also agree with more subtle aspects. The ratio of upper
and lower caustic populations, the velocity shift at the tip of the
wedge from M31's systemic velocity, and the slightly skewed shape of the
wedge all have physical
meaning in the simulations, and the data are in good agreement with
the simulation trends for all these aspects.  The high
precision of the kinematic data encourages use in constraining our
satellite model further.  For example, we already had to adjust the
radii in the model very slightly, an adjustment well within the
constraints posed by other observations of the stream debris.

Together, the metallicity and kinematics make a strong case that the W
Shelf indeed originates from the same disruption event that produced
the GSS.  The structure around M31 is still incompletely understood.
In particular, part of the GSS region shows a second stream offset 
by $+100 \kms$ from the GSS itself, which is currently unexplained by our
models.  Is it possible that the W Shelf itself results from a
similar, but independent accretion event to that which formed the GSS?
The rich variety of observational features matched by our simulations
suggests that this is unlikely.  However, further confirmation of the
model is worthwhile.  The next test for the model is likely to come in
the NE Shelf region, where the model places the bulk of the
GSS-related material.  In this region our model predicts a similar
wedge kinematic pattern to that in the W Shelf (see F07).  We have
argued that negative-velocity (outflowing) stars have been detected
already in this region, but the detection of a positive-velocity
(returning) component will provide a clear connection between all
parts of the GSS model.  
The progenitor's total stellar mass in our model is roughly comparable
to that of the LMC, and adding up the stellar mass of all the
GSS-related features will provide another test of the model.  

The smooth spheroid component is consistent with that found elsewhere
in M31.  The velocity distribution agrees in its mean and dispersion
with that found previously on the opposite side of M31's disk, as does
the metallicity distribution.  This provides strong evidence for a
symmetric, mostly virialized hot component at intermediate radii.
It may be a bit simplistic to call this inner spheroid ``smooth'',
as evidenced by the $-175 \kms$ clump of stars with a 
small range of metallicities we found in \S~\ref{sec.analysis}.  
This is not unexpected in
theories of galaxy halo formation, as fine-grained structure can
linger in phase space and stellar population space far longer than is
apparent in morphology \citep[e.g.][]{helmi99,johnston08}.  However,
at a minimum it is made up of a very large number of components
below the threshold of individual detectability spread over a large
velocity range, as opposed to the cold monolithic kinematic pattern
of the shelf component.

It is clear that M31's inner halo remains a confusing region, and much
work remains to be done to decompose it into kinematically distinct
components.  The extent of the W Shelf in particular remains to be
mapped, as in star-count maps it becomes confused with M31's outer
disk on both the NW and SE sides.  
Extrapolating from progress in the last decade, we expect rapid growth
in our understanding of M31's halo from the combination of large
imaging surveys with spectroscopic surveys such as we have presented
here.

The wedge pattern detected in Fig.~\ref{fig.rvcomp} is to our
knowledge the first time a radial shell kinematic pattern 
\citep{merrifield98} has been
clearly detected in any galaxy, although features of the
pattern were suggested earlier in F07.  It is also the first spiral
galaxy found to exhibit a system of multiple shell structures, though
recently NGC 4651 has been found to exhibit similar structures
\citep{dmd10}.  Shell systems are of course common in elliptical
galaxies \citep[e.g.][]{malin83,merrifield98,canalizo07}.  Our results
offer inspiration for the future detection of shell kinematics in more
distant galaxies and its use in measuring their gravitational
potential, using either bright tracers such as planetary nebulae or
globulars or else RGB stars detected with a future generation of
telescopes.  

\section*{ACKNOWLEDGMENTS}
We are grateful to Evan Kirby for his 
substantial assistance with the mask reduction.
We thank Tom Quinn and Joachim Stadel for the use
of \softfont{PKDGRAV}, and Josh Barnes for the use of \softfont{ZENO}.
We also thank Tom Brown for sending the surface brightness 
values used in normalizing our spheroid model,
Mike Irwin for providing the INT star-count map 
used in Fig.~\ref{fig.masks},
and Marla Geha for helpful discussions.
The data presented herein were obtained at the W.M. Keck Observatory,
which is operated as a scientific partnership among the California
Institute of Technology, the University of California and the National
Aeronautics and Space Administration. The Observatory was made
possible by the generous financial support of the W.M. Keck Foundation.
The authors wish to recognize and acknowledge the very significant
cultural role and reverence that the summit of Mauna Kea has always
had within the indigenous Hawaiian community.  We are most fortunate
to have the opportunity to conduct observations from this mountain.
MF acknowledges support from NSF grant AST-1009652,
PG was supported by AST-1010039 and NASA grant HST-GO-12105.03 from STScI.
KG was supported by NASA through Hubble Fellowship 51273.01 from STScI.
EJT was supported by a Graduate Assistance in Areas of National Need (GAANN) Fellowship.
MC was supported by a Grant-in-Aid for Scientific
Research (20340039) of the Ministry of Education, Culture,
Sports, Science and Technology in Japan.

{\footnotesize
\bibliographystyle{mn2e}
\bibliography{m31}
}

\appendix
\section{Physics of less-idealized shells --- expansion, rotation,
and velocity dispersion} 
Here we derive some results on the kinematics of non-ideal shells, for
use both in the preceding analysis of the W Shelf and in understanding
future observations of other stellar shells.

\citet{merrifield98} derived the observable kinematics of a stellar
shell in the idealized case of a monoenergetic, perfectly radial flow
of stars in a gravitational field of uniform strength.  In the same
approximation, F07 derived the shell surface density in terms of the
gravitational potential and flow rate of stars through the shell.
Both papers discussed several physical and observational aspects 
of stellar shells besides those mentioned in this paper.  However, 
here we need several results not mentioned in either paper.

To establish notation, we repeat the argument of \citet{merrifield98}
here.  Denote the three-dimensional radius as $r$ and the projected
radius as $R$.  Consider a shell of fixed radius $\rsh$, with stars of
constant orbital energy flowing out to $\rsh$ and back, and for convenience
let us consider only the stars on the far side of the host. Write 
$b = r/\rsh$, $\beta = R/\rsh$.  Let the gravitational field have a
constant strength $g$; then at the shell edge the circular velocity is
$V_c = (g \rsh)\half$.  
The radial velocity of the stars is given by
$v_r^2 = 2 g (\rsh - r) = 2 V_c^2 (1-b)$.  
The line-of-sight velocity is instead given by 
$\vlos = (z/r) v_r$, where $z$ is the distance relative to the 
host distance.  (Here we neglect perspective effects.)  Near the shell
edge we can to first order take 
$z/r = [(r^2-R^2)/r^2]\half \approx [2 (b-\beta)]\half$.
Then 
$\vlos = \pm 2 V_c (b-\beta)\half (1-b)\half$.  The magnitude of
the velocity is maximized at 
$b = (1+\beta)/2$, implying the envelope of $\vlos$ is given by
$v_{env} = \pm V_c (1-\beta) = \pm V_c (\rsh - R)/\rsh$.  Here the top and
bottom signs apply to the outgoing and incoming stars
respectively. 
Thus projecting the velocities has squashed the sideways parabola of
$v_r(r)$ into a wedge.  
The gradual turnover in $\vlos$ around its maximum gives
the distribution of stars in velocity space a caustic or horn shape.
A similar argument applies to stars on the
near side of the host, with the roles of the positive and negative
caustics reversed.

While undoubtedly useful as a first approximation, this contains one
fundamental problem: the radial shell exists only because stars have
different orbital energies and hence arrive at the shell edge at
different times.  The gradient is invariably in the sense that
later-arriving stars have higher energies, because larger orbital
energy increases the orbital period.  We can handle this problem
simply by incrementing the outward velocity of all stars in the shell
by the same amount.
\begin{equation}
(v_r-\vsh)^2 = 2 g (\rsh - r) = 2 V_c^2 (1-b)
\end{equation}
If we label each star by the time $t_p$ it arrives at the shell boundary,
its specific energy becomes $E = \vsh^2/2 + \rsh g + \vsh g t_p$.  The energy
gradient can then be expressed as 
\begin{equation}
dE/d(t_p) = \vsh g = \vsh V_c^2 / \rsh\; . 
\end{equation}
(As argued above, $dE/d(t_p)$ and $\vsh$ are both in general positive.)
This shows the expansion speed of the shell is proportional to the
gradient in orbital energy along the stream.

We can again solve for the maximum line-of-sight velocity at a given
projected radius.  Defining 
$\epsilon \equiv \vsh / (g \rsh)\half = \vsh / V_c$,
$\delta = 1-\beta$, and 
$\nu = (\epsilon/8) ([16\delta+\epsilon^2]\half - \epsilon)$,
and differentiating to find the extremum,
we find an envelope velocity of
\begin{equation}
\label{eqn.expcaustic}
v_{env} = V_c (\delta \pm \nu)\half 
[\epsilon \pm (\delta \mp \nu)\half] \; .
\end{equation}
This results in an envelope in $r$-$v_{los}$ space that is shifted
upwards relative to the monoenergetic case at most radii, with an
increasingly skewed shape approaching the shelf radius so that the
tip remains at zero velocity.  This is a natural result of the
outgoing stars (upper branch of the envelope) having higher energies
and thus greater velocities.  
For lines of sight outside the radius where the ``returning'' particles have zero
radial velocity, given by $(2\delta)\half = \epsilon$, there is
no returning caustic and the minimum envelope velocity of zero
is reached at the midplane.
The tip remains fixed at zero velocity
because the radial velocity at the shelf edge lies perpendicular to
the line of sight.  If the shell material instead lies on the near
side of the host galaxy, then the energy gradient instead skews the
envelope in the negative-velocity direction, with the tip again 
fixed at zero.  If
the shell material fills both sides of the galaxy, the degeneracy of
the near and far side envelopes is broken and both envelopes will be
present, offset from each other at all $r < r_{sh}$.

Since the expansion velocity is typically small compared to $V_c$, 
it is more useful to take a perturbative approach, which allows
us to incorporate second-order effects in the gravitational potential
and velocity projection.  To this order we can write 
$z/r \approx \sqrt{2} (\delta-D)\half (1 + \frac{3}{4}D - \frac{1}{4} \delta)$.
with $D \equiv 1-b$.
We assume a flat rotation curve, in which case 
$\Phi(r)-\Phi(\rsh) \approx -V_c^2 D (1 + \frac{1}{2} D)$.
To this order we can assume the extremum is reached at $D = \delta/2$,
the lowest-order solution.  Combining the various factors, we reach our
final approximation
\begin{equation}
\label{eqn.causticapprox}
v_{env} \approx V_c \delta\half (1 + \frac{1}{8} \delta)
  \left[ \epsilon \pm \delta\half (1 + \frac{1}{8} \delta) \right]      \; .
\end{equation}
The first factor of $(1 + \frac{1}{8} \delta)$ is contributed by the
projection factor, and the second (in the case of a flat rotation curve only)
by the potential.  

In the case of the W Shelf, as will be true in most cases, the shell
material is not purely radial but instead has significant angular
momentum.  This affects the observed velocities by shifting the
envelope up or down.  We
apply this as a global shift to the entire envelope in
Fig.~\ref{fig.rvcomp}.  A higher-order analysis would take into
account the radial dependence of tangential velocity and derive the
envelope shift by rederiving the caustic analysis above, but the simple
shift used here appears accurate enough for our purposes.

As a final exercise, we estimate the smearing in the stellar velocities
near the caustic edge induced by a local radial velocity dispersion $\sigma_r$
of the stream particles.  This dispersion might be left over from initial
conditions in a young stream like that discussed here, or else result 
from interaction with LCDM substructure \citep{carlberg09} in an older one.
To first order we can estimate the observed dispersion from the
monoenergetic picture, where the 
projection factor at the velocity caustic is $z/r = (1-\beta)\half$.  
Thus we expect the envelope edge to be blurred by an amount 
\begin{equation}
\label{eqn.causticsmear}
\sigma_{los} = [(\rsh-R)/\rsh)]\half \sigma_r \; .
\end{equation}

\label{lastpage}
\end{document}